\documentclass[aps,preprintnumbers,superscriptaddress,showpacs]{revtex4}
\usepackage{epsfig}
\usepackage{psfrag}
\usepackage{amsfonts}
\usepackage{graphicx}
\usepackage{dcolumn}
\usepackage{float}
\usepackage{bm}

\begin{document}

\title{Entropic destruction of heavy quarkonium with hyperscaling
violation}

\author{Ang Li}
\affiliation{School of Mathematics and Physics, China University
of Geosciences, Wuhan 430074, China}

\author{Zi-qiang Zhang}
\email{zhangzq@cug.edu.cn} \affiliation{School of Mathematics and
Physics, China University of Geosciences, Wuhan 430074,
China}\affiliation{Key Laboratory of Quark and Lepton Physics
(MOE), Central China Normal University, Wuhan 430079, China}

%%%%%%%%%%%%%%%%%%%%%%%%%%%%%%%%%%%%%%%%
\begin{abstract}
We study the entropic destruction of heavy quarkonium in strongly
coupled theories with an anisotropic scaling symmetry in time and
spatial direction. We consider Lifshitz and hyperscaling violation
theories which are covariant under a generalized Lifshitz scaling
symmetry with the dynamical exponent $z$ and hyperscaling
violation exponent $\theta$. It is shown that the entropic force
depends on the parameters of these theories. In particular,
increasing $z$ decreases the entropic force thus reducing the
quarkonium dissociation, while increasing $\theta$ has opposite
effect.
\end{abstract}

\maketitle
%%%%%%%%%%%%%%%%%%%%%%%%%%%%%%%%%%%%%%%%
\section{Introduction}
It is believed that the heavy ion collisions at Relativistic Heavy
Ion Collider (RHIC) and Large Hadron Collider (LHC) have produced
a new state of matter so-called strongly coupled quark gluon
plasma (QGP) \cite{s1,s2,s3}. One of the signatures of QGP
formation is the dissociation of heavy quarkonium \cite{s4}.
Previous research has indicated that the immediate cause of this
dissociation is Debye screening. However, there is a puzzle
revealed in the recent experiment of charmonium($c\bar{c}$): the
$c\bar{c}$ suppression at RHIC (lower energy density) was stronger
than that at LHC (larger energy density) \cite{s5,s6}. Obviously,
this finding is in conflict with the Debye screening and the
thermal activation \cite{s7,s8}. To explain this, D. E Kharzeev
argued \cite{s9} that an anomalously strong suppression of
$c\bar{c}$ near the deconfinement transition could be a
consequence of the nature of deconfinement. This inference was
based on the lattice QCD results \cite{s10,s11,s12,s13} which
indicated a large amount of entropy $S$ associated with the heavy
quarkonium (or $q\bar{q}$) in QGP and this entropy was found to
grow as a function of the distance between $q\bar{q}$. In the
proposal of \cite{s9}, the entropy gives rise to the entropic
force
\begin{equation}
\mathcal{F}=T\frac{\partial S}{\partial L},\label{f}
\end{equation}
where $T$ is the temperature of the plasma and $L$ is the
inter-quark distance of $q\bar{q}$. It has been conjectured
\cite{s9} that the entropic force is the leading role in the
deconfinement transition itself and a possible relation of
the observed peak in the entropy near the deconfinement transition
to the long string condensation
\cite{s14,s15,s16,s17,s18,s19,s20,s21,s22}.

AdS/CFT correspondence \cite{s23,s24,s25}, which maps a $d$
dimensional quantum field theory to its dual gravitational theory,
living in $d+1$ dimensional, has yielded many important insights
for studying different aspects of QGP \cite{JCA}. Using AdS/CFT,
D. Kharzeev first calculated the entropic force for
$\mathcal{N}=4$ super Yang-Mills (SYM) plasma \cite{KHA}. Therein,
it was found that the narrow and strong peak in the entropy near
the transition temperature is related to the deconfinement of
quarkonium. Soon after, this idea was extended to various cases,
e.g., the entropic force of moving quarkonium was investigated in
\cite{KBF}. The chemical potential effect on the entropic force
was studied in \cite{ZQ}. Also, this quantity has been addressed
from AdS/QCD \cite{II}. Other related results can be found in
\cite{DE1,ZQ1,ZQ2}.

In this paper, we are interested in studying the entropic force in
strongly coupled theories with an anisotropic scaling symmetry in
time and spatial direction. In particular, we will consider
hyperscaling violation theories \cite{s26,s27,s28,s29,s30} which
are covariant under a generalized Lifshitz scaling symmetry with
the dynamical exponent $z$ and hyperscaling violation exponent
$\theta$ and apply their gravity duals. Because of different
scaling of time and space, these theories are intrinsically
non-relativistic. On the other hand, the metrics of these theories
are scale invariant but not conformally invariant. From a QCD
point of view, such theories could be used as a basis for new
AdS/QCD constructions. Recently, the hyperscaling violation
theories have been used to describe the string theory
\cite{s31,s32,s33,s34,s35} and holographic superconductors
\cite{s36,s37,s38,s40}. Also, some QCD observables or quantities,
e.g., Schwinger effect \cite{s42}, heavy quark potential
\cite{s48,s49} have been addressed in such theories. Motivated by
this, in this paper, we study the entropy force in hyperscaling
violation theories. Specifically, we want to understand how the
non-relativistic parameters $z$ and $\theta$ modify the entropy
force as well as the quarkonium dissociation.

The paper is organized as follows. In the next section, we
introduce the hyperscaling violation theories given in \cite{s35}.
In section 3, we study the behavior of the entropic force with
respect to a heavy quarkonium in these theories and analyze the
effects of $z$ and $\theta$ on it. In section 4, we summarize the
results and make some discussions.

\section{Hyperscaling Violation background}
Let's start with a brief review of the backgrounds with
hyperscaling violation \cite{s35}. The corresponding metric takes
the form \cite{cc,bg}
\begin{equation}
ds^2=u^{\theta}[-\frac{dt^2}{u^{2z}}+\frac{b_0du^2+dx^idx^j}{u^2}], \label{metric0}
\end{equation}
and is invariant under a generalized Lifshitz scaling $t\to
\Lambda^zt,u\to\Lambda u,x^i\to\Lambda x^i,ds^2\to
\Lambda^{-\theta }ds^2$, where $b_0=\ell^2$ with $\ell$ the IR
scale. $z$ is the dynamical Lifshitz parameter (or is called the
dynamical critical exponent) which characterizes the behavior of
system near the phase transition. $\theta$ denotes the
hyperscaling violation exponent which is responsible for the
nonstand scaling of physical quantities and controls the
transformation of the metric.

The scalar curvature is
\begin{equation}
\mathcal{R}=-\frac{3\theta^2-4(z+3)\theta+2(z^2+3z+6)}{b_0}u^{-\theta}.
\end{equation}

Note that the geometry is flat for $\theta=2$, $z=0,1$. It is
Ricci flat for $\theta=4$, $z=3$, and in Ridler coordinates for
$\theta=0$, $z=1$.

Using a radical redefinition
\begin{equation}
 u=(2-z)r^{1/(2-z)},
 \end{equation}
and rescaling $x^i$ and $t$, one gets:
 \begin{equation}
 ds^2\sim r^{\frac{\theta-2}{2-z}}[-f(r){dt^2}+\frac{dr^2}{f(r)}+dx^idx^i]\label{sm},\qquad
 f(r)=f_0(\frac{r}{\ell})^{\frac{2(1-z)}{2-z}},\label{key}
 \end{equation}
with $f_0=(2-z)^{2(1-z)}$. The boundary is at $r=\infty$ when
$\frac{\theta-2}{2-z}<0$ and at $r=0$ when
$\frac{\theta-2}{2-z}>0$.

The energy scale reads
 \begin{equation}
 E\simeq u^{(\theta -2z)/2}\simeq r^{(\theta -2z)/2(2-z)}.
 \end{equation}

As discussed in \cite{s35}, the metrics (\ref{sm}) satisfy the
Gubser conditions (in the context of holography, many non-trivial
flows become singular upon dimensional reduction, leading to the
Gubser criterion for the acceptability of a naked singularity
\cite{ss}), given by
 \begin{equation}
 \frac{2z+3(2-\theta)}{2(z-1)-\theta}>0,\qquad
 \frac{z-1}{2(z-1)-\theta}>0,\qquad
 \frac{2(z-1)+3(2-\theta)}{2(z-1)-\theta}>0,\label{7}
 \end{equation}
and the thermodynamic stability condition
 \begin{equation}
 \frac{z}{2(z-1)-\theta}>0. \label{8}
 \end{equation}

The generalizations of (\ref{sm}) to include finite temperature
are
\begin{equation}
 ds^2\sim(\frac{r}{\ell})^{-\alpha}[-f(r){dt^2}+\frac{dr^2}{f(r)}+dx^idx^i],\label{L1}
\end{equation}
with
\begin{equation}
f(r)=f_0(\frac{r}{\ell})^{2\beta}[1-(\frac{r}{r_h})^\gamma],
\end{equation}
where
\begin{equation}
\alpha=\frac{\theta-2}{z-2}, \qquad \beta=\frac{z-1}{z-2}, \qquad
\gamma=\frac{2z+3(2-\theta)}{2(2-z)}.
\end{equation}

The Hawking temperature reads
\begin{equation}
T=\frac{f_0}{8\pi\ell}\left(\frac{r_h}{\ell}\right)^{\frac{z}{z-2}}|\frac{2z-3\theta+6}{z-2}|.\label{10}
\end{equation}

For more about the hyperscaling violation backgrounds, we refer to \cite{s35}.

\section{Entropic force}
We now proceed to holographically compute the entropic force for
the background metric (\ref{L1}) following \cite{KHA}. The
Nambu-Goto action is
\begin{equation}
S_{NG}=T_F\int d\tau d\sigma\mathcal
L=-\frac{1}{2\pi\alpha^\prime}\int d\tau d\sigma\sqrt{-det
g_{\alpha \beta}},
 \qquad  \label{S}
\end{equation}
where $T_F=\frac{1}{2\pi\alpha^\prime}$ is the fundamental string
tension and $\alpha^{\prime}$ is related to the 't Hooft coupling
$\lambda$ by $\frac{R^2}{\alpha^\prime}=\sqrt{\lambda}$ (hereafter
we set $R=1$). $g$ represents the determinant of the induced
metric with
\begin{equation}
g_{\alpha\beta}=g_{\mu\nu}\frac{\partial
X^\mu}{\partial\sigma^\alpha} \frac{\partial
X^\nu}{\partial\sigma^\beta},
\end{equation}
where $X^\mu$ and $g_{\mu\nu}$ are the target space coordinates
and metric, respectively.

For our purpose, we take the static gauge
\begin{equation}
t=\tau, \qquad x=\sigma, \qquad y=0,\qquad z=0, \label{par}
\end{equation}
and assume $r$ depends only on $\sigma$
\begin{equation}
r=r(\sigma),
\end{equation}
given that, the induced metric becomes
\begin{equation} g_{00}=-b(r)f(r), \qquad g_{01}=g_{10}=0, \qquad
g_{11}=b(r)[1+\frac{\dot {r}}{f(r)})],
\end{equation}
with $b(r)=(r/\ell)^{-\alpha}$, $\dot r=dr/d\sigma$, then the
Lagrangian density reads
\begin{equation}
\mathcal L=\sqrt{b^2(r)[f(r)+\dot r^2]}.\label{ll}
\end{equation}

Since $\mathcal L$ does not depend on $\sigma$ explicitly, the
corresponding Hamiltonian is a constant, that is
\begin{equation}
\mathcal H=\frac{\partial\mathcal L}{\partial
\dot{r}}\dot{r}-\mathcal L=Constant.
\end{equation}

Imposing the boundary condition at the tip of the U-shaped string
\begin{equation}
\dot r=0,\qquad  r=r_c\qquad (r_h>r_c)\label{con},
\end{equation}
one gets
\begin{equation}
\dot{r}=\sqrt{\frac{f^2(r)b^2(r)-f(r)f(r_c)b^2(r_c)}{f(r_c)b^2(r_c)}}\label{kk},
\end{equation}
with
\begin{equation}
f(r_c)=f_0(\frac{r_c}{\ell})^{2\beta}
[1-(\frac{r_c}{r_h})^\gamma],\qquad
b(r_c)=(\frac{r_c}{\ell})^{-\alpha}.
\end{equation}

The inter-distance of $q\bar{q}$ can be calculated by integrating
(\ref{kk})
\begin{equation}
L=2\int_0^{r_c}dr\sqrt{\frac{A(r_c)B(r)}{A^2(r)-A(r)A(r_c)}}\label{dotr}.
\end{equation}
with
\begin{equation}
A(r)=b^2(r)f(r),\qquad B(r)=b^2(r),\qquad A(r_c)=b^2(r_c)f(r_c),
\end{equation}
the shape of $L$ against $\varepsilon$ with $\varepsilon=r_c/r_h$
is shown in Fig. 1. Here we take $z=1.6,\theta=1$ (other values of
$z$ and $\theta$ lead to similar picture).

\begin{figure}[H]
\centering
\includegraphics[width=8cm]{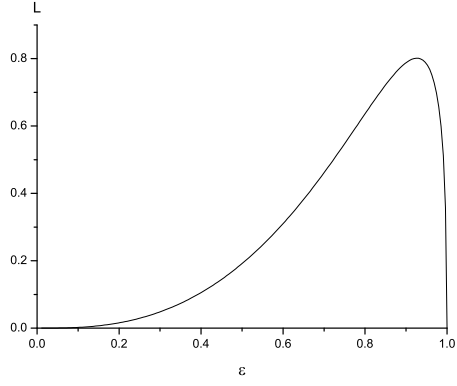}
\caption{$L$ versus $\varepsilon$. Here $z=1.6,\theta=1$.}
 \end{figure}

The next task is to calculate entropy, given by
\begin{equation}
S=-\frac{\partial F}{\partial T},\label{free}
\end{equation}
where $F$ is the free energy of the $q\bar{q}$, which is related
to the on-shell action of the fundamental string in the dual
geometry. Note that this quantity has been holographically studied
at zero temperature \cite{jm} and finite temperature \cite{ab,sj}.
Broadly speaking, there are two cases for $F$.

1. If $L>\frac{c}{T}$ ($c$ denotes the maximum value of $LT$),
some new configurations should be considered \cite{db} and then
there are a few possible choices of $F$ \cite{mc}, e.g., one may
choose a configuration of two disconnected trailing drag strings
\cite{s44,s45}. Under such conditions, the free energy is
\begin{equation}
F^{(1)}=\frac{1}{\pi \alpha^{\prime}}\int_0^{r_c}dr,
\end{equation}
leading to
\begin{equation}
S^{(1)}=\sqrt{\lambda}\theta(L-\frac{c}{T}),
\end{equation}
where $\theta (L-\frac{c}{T})$ is the Heaviside step function.

2. If $L<\frac{c}{T}$, the fundamental string is connected, then
the free energy of the $q\bar{q}$ could be obtained by plugging
(\ref{kk}) into (\ref{ll}), that is
\begin{equation}
F^{(2)}=\frac{1}{\pi\alpha^{\prime}}\int_0^{r_c}\sqrt{\frac{A(r)B(r)}{A(r)-A(r_c)}}\label{dotr},
\end{equation}
then from (\ref{10}), (\ref{free}) and (\ref{dotr}), one gets
\begin{equation}
S^{(2)}=-\frac{\partial F^{(2)}}{\partial T}=-\frac{\partial
F^{(2)}}{\partial r_h}/\frac{\partial T}{\partial r_h},\label{ff}
\end{equation}
where
\begin{equation}
\frac{\partial F^{(2)}}{\partial
r_h}=\frac{1}{2\pi\alpha^{\prime}}\int_0^{r_c}\frac{A^\prime(r)B(r)[A(r)-A(r_c)]-A(r)B(r)[A^{\prime}(r)-A^{\prime}(r_c)]}{\sqrt{A(r)B(r)[A(r)-A(r_c)]^3}},
\end{equation}
\begin{equation}
\frac{\partial T}{\partial
r_h}=\frac{f_0}{8\pi\ell}\left|\frac{2z-3\theta+6}{z-2}\right|\ell^{\frac{z-2}{z}}\frac{z}{z-2}r_h^{\frac{2}{z-2}},
\end{equation}
with
\begin{eqnarray}
A^{\prime}(r)&=&b^2(r)f_0(\frac{r}{\ell})^{2\beta}\gamma r^\gamma r_h^{-\gamma-1},\\
A^{\prime}(r_c)&=&A^{\prime}(r)|_{r=r_c},
\end{eqnarray}
note that by plugging $z=1$ and $\theta=0$ in (\ref{ff}),  the
entropic force for $\mathcal{N}=4$ SYM \cite{KHA} can be
reproduced, as expected.

Next, we calculate the entropy force. Before we get to that, we
determine the value range of $z$ and $\theta$. According to the
Gubser condition of (\ref{7}) and the thermodynamic stability of
(\ref{8}), one has
\begin{equation}
1<z<2, \qquad \theta<2, \qquad \theta<z.
\end{equation}

Taken together, one can numerically calculate the entropic force
from (\ref{ff}). In Fig.2, we plot $S^{(2)}/\sqrt{\lambda}$ versus
$LT$ for different values of $z$ and $\theta$, here we have used
the relation $\frac{1}{\alpha^\prime}=\sqrt{\lambda}$. From the
left panel, one can see that with fixed $z$, increasing $\theta$
leads to larger entropy at small distances. While from the right
panel one finds that with fixed $\theta$, increasing $z$ leads to
smaller entropy at small distances. As stated earlier, the
entropic force is found to grow as a function of the distance (see
(\ref{f})) and responsible for the quarkonium dissociation. Thus,
one concludes that increases $\theta$ increases the entropic force
thus enhancing the quarkonium dissociation, while $z$ has opposite
effect.

To analyze how the temperature influences the entropic force in
the hyperscaling violation backgrounds, we plot
$S^{(2)}/\sqrt{\lambda}$ versus $L$ with fixed $z$ and $\theta$ in
Fig.3. One finds increasing $T$ leads to increasing the entropic
force, in accord with the finding in \cite{s46}. The physical
meaning of our results will be discussed in the next section.

\begin{figure}[H]
\centering
\includegraphics[width=8cm]{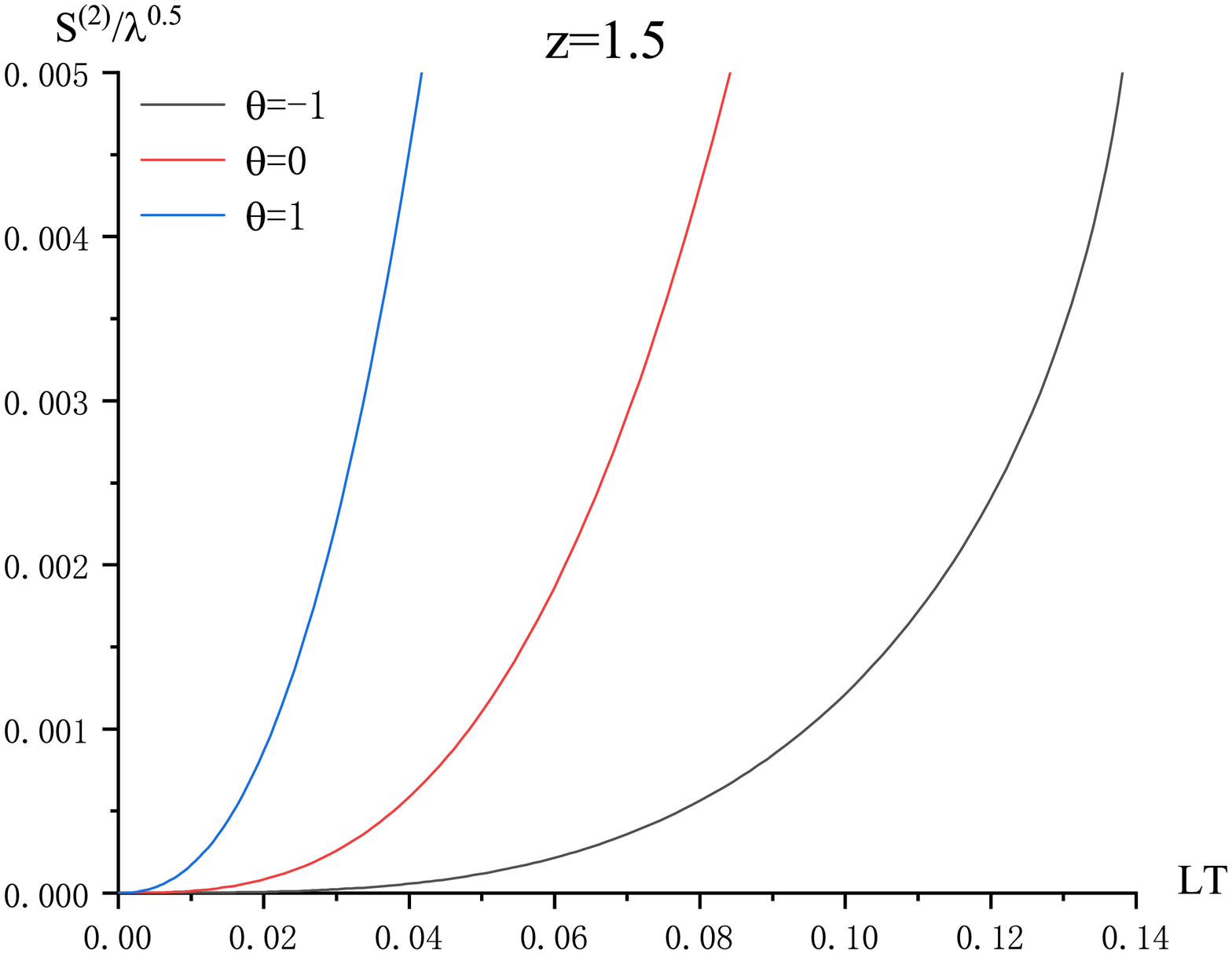}
\includegraphics[width=8cm]{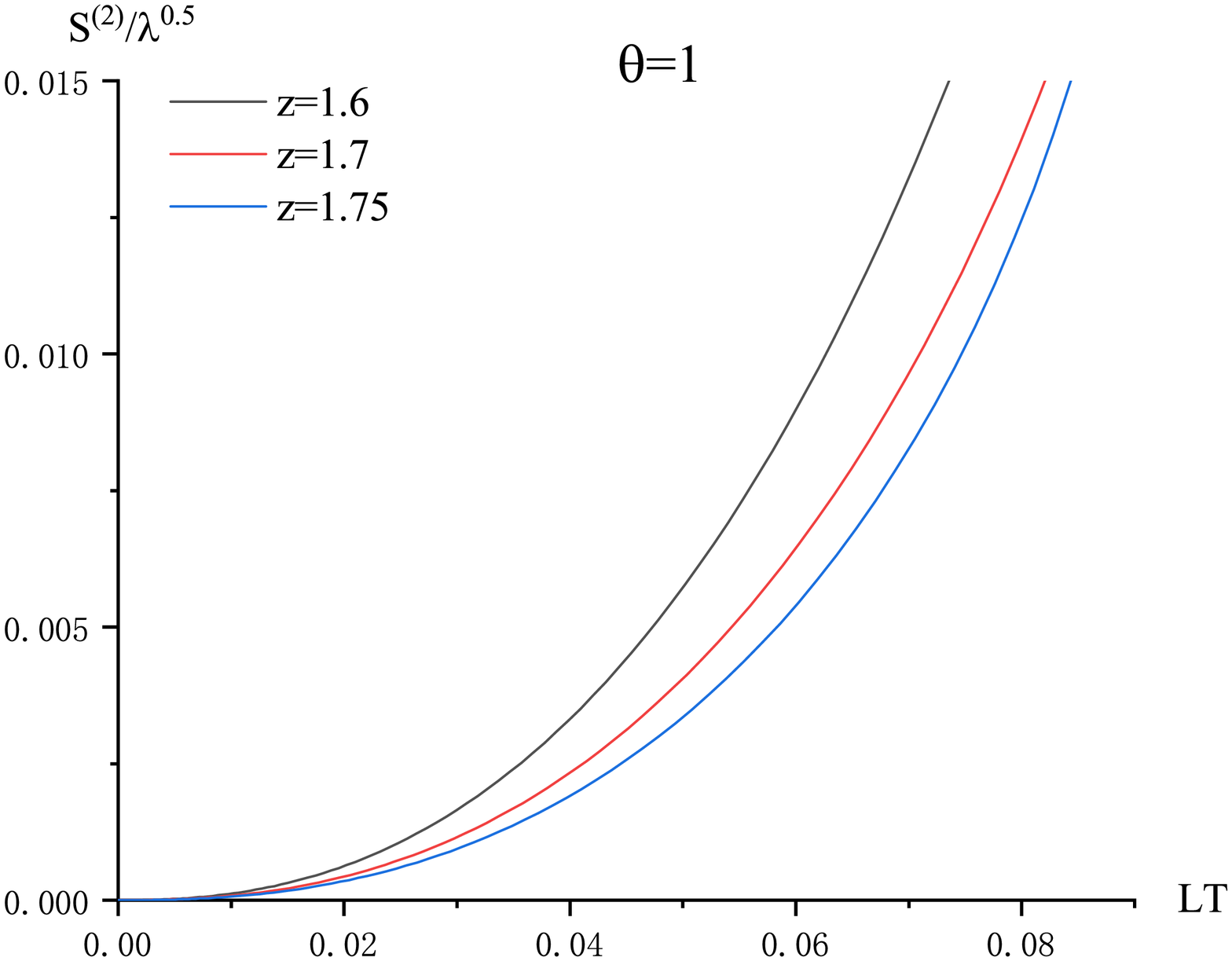}
\caption{$S^{(2)}/\sqrt{\lambda}$ versus $LT$ for different
$\theta$ and $z$. Left: $z=1.5$ and from left to right
$\theta=1,0,-1$. Right: $\theta=1$ and from left to right
$z=1.6,1.7,1.75$.}
\end{figure}

\begin{figure}[H]
\centering
\includegraphics[width=8cm]{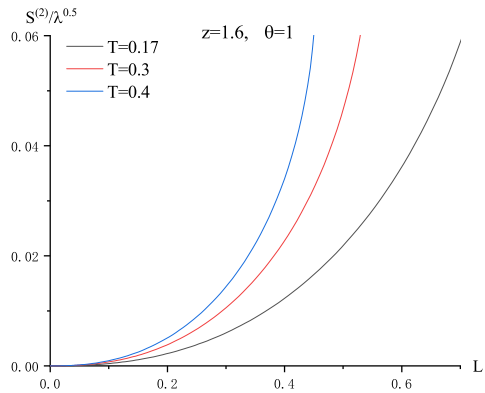}
\caption{$S^{(2)}/\sqrt{\lambda}$ versus $L$. Here
$z=1.6,\theta=1$, and from top to bottom $T=0.4,0.3,0.17$,
respectively.}
\end{figure}

\section{Conclusion and Discussion}

In this paper, we studied the entropic destruction of heavy
quarkonium in the non-relativistic backgrounds. We considered
Lifshitz and hyperscaling violation theories which are strongly
coupled theories with an anisotropic scaling symmetry in the time
and a spatial direction. An understanding of how the entropic
force changes by these theories may be essential for theoretical
predictions. We discussed how the entropic force depends on the
non-relativistic parameters $z$ and $\theta$. It is shown that
increasing $z$ decreases the entropic force thus reducing the
quarkonium dissociation. Namely, the quarkonium melts harder in
the presence of an anisotropic scaling symmetry in the time and a
spatial direction. On the other hand, we find that increases
$\theta$ increases the entropic force, indicating the inclusion of
$\theta$ (means hyperscaling violation in the dual field theory)
enhances the quarkonium dissociation.

Here we would like to point out that the heavy quarkonium
discussed here mainly refers to charmonium. Because most of the
bottomonium ($b\bar{b}$) has smaller sizes, and is thus much less
affected by the entropic force \cite{DE1}. Actually, the available
data shows that the $b\bar{b}$ suppression is indeed stronger at
LHC \cite{sc,bb} than at RHIC \cite{la,aa}.

Finally, it is relevant to mention that the imaginary potential
\cite{jno} has been proposed to responsible for melting the heavy
quarkonium recently. It would be interesting to investigate this
quantity in hyperscaling violation theories as well. We leave this
for further study.

\section{Acknowledgments}
This work is supported by the Innovation Fund of of Key Laboratory
of Quark and Leption Physics under grant No. QLPL2020P01.

\end{document}